\documentclass[conference]{IEEEtran}
\usepackage{amssymb}

\usepackage{amsmath,fancyhdr,graphicx,amsthm,amsfonts,epsfig,amssymb,color}
\usepackage{algorithmic}
\usepackage{algorithm}

\newcommand{\R}{\mathbb{R}}

\renewcommand{\P}{\mathcal{P}}

\renewcommand{\mathbf}{\boldsymbol}

\newtheorem{theorem}{Theorem}
\newtheorem{lemma}[theorem]{Lemma}
\newtheorem{prop}[theorem]{Proposition}

\renewcommand{\Re}{\mathbb{R}}
\newcommand{\<}{\left\langle}
\renewcommand{\>}{\right\rangle}

\begin{document}
\title{Stable Principal Component Pursuit}


%
\author{\IEEEauthorblockN{Zihan Zhou\IEEEauthorrefmark{1},
Xiaodong Li\IEEEauthorrefmark{2}, John
Wright\IEEEauthorrefmark{3}, Emmanuel
Cand{\`e}s\IEEEauthorrefmark{2}\IEEEauthorrefmark{4} and Yi
Ma\IEEEauthorrefmark{1}\IEEEauthorrefmark{3}}
\IEEEauthorblockA{\IEEEauthorrefmark{1}Electrical and Computer
Engineering, UIUC, Urbana, IL 61801}
\IEEEauthorblockA{\IEEEauthorrefmark{2}Department of
Mathematics, Stanford University, Stanford, CA 94305}
\IEEEauthorblockA{\IEEEauthorrefmark{3}Microsoft Research Asia,
Beijing, China}
\IEEEauthorblockA{\IEEEauthorrefmark{4}Department of
Statistics, Stanford University, Stanford, CA 94305}}

\maketitle
\begin{abstract}
In this paper, we study the problem of recovering a low-rank
matrix (the principal components) from a high-dimensional data
matrix despite both small entry-wise noise and gross sparse
errors. Recently, it has been shown that a convex program,
named Principal Component Pursuit (PCP), can recover the
low-rank matrix when the data matrix is corrupted by gross
sparse errors. We further prove that the solution to a related
convex program (a relaxed PCP) gives an estimate of the
low-rank matrix that is simultaneously stable to small
entry-wise noise and robust to gross sparse errors. More
precisely, our result shows that the proposed convex program
recovers the low-rank matrix even though a positive fraction of
its entries are arbitrarily corrupted, with an error bound
proportional to the noise level. We present simulation results
to support our result and demonstrate that the new convex
program accurately recovers the principal components (the
low-rank matrix)  under quite broad conditions. To our
knowledge, this is the first result that shows the classical
Principal Component Analysis (PCA), optimal for small i.i.d.
noise, can be made robust to gross sparse errors; or the first
that shows the newly proposed PCP can be made stable to small
entry-wise perturbations.
\end{abstract}

\section{Introduction}
The advance of modern information technologies has produced
tremendous amount of high-dimensional data in science,
engineering, and society, such as images, videos, web
documents, and bioinformatics data. It has become a pressing
challenge to develop efficient and effective tools to process,
analyze, and extract useful information from such
high-dimensional data. One of the fundamental problems here is
how to extract the intrinsic low-dimensional structure of such
high-dimensional data.

\paragraph{Classical Principal Component Analysis}
Arguably, the classical {\em Principal Component Analysis}
(PCA) \cite{Eckart36, Jolliffe} is the most widely used
statistical tool for high-dimensional data analysis and
dimensionality reduction today. It basically assumes that the
data approximately lie on a low-dimensional linear subspace.
Mathematically, if we stack all the data points as column
vectors of a matrix $M$, then the matrix should be
approximately low-rank and can be written as $ M = L_0 + Z_0$,
where $L_0$ is a low-rank matrix (representing the subspace)
and $Z_0$ models a small noisy perturbation of each entry of
$L_0$. Then, PCA simply seeks the best rank-$k$ estimate of
$L_0$ in the $\ell_2$ sense, which can be solved efficiently
via singular value decomposition (SVD) and thresholding. It can
be shown that if the perturbation is i.i.d. Gaussian, this
gives a statistically optimal estimate of the subspace. Such an
estimate is naturally stable in the sense that the error is
bounded to be proportional to the magnitude of the
perturbation.

\paragraph{Robust PCA via Principal Component Pursuit}
However, it is well known that the classical PCA breaks down
even with a single grossly corrupted entry in the data matrix
$M$, i.e., it is {\em not robust} to gross errors or outliers.
Many methods have been proposed to alleviate this problem,
however, none of them yield a polynomial-time algorithm with
strong performance guarantees (see \cite{Wright2009-pp} for a
detailed discussion).




The recently proposed Principal Component Pursuit (PCP) method
utilizes a convex program that guarantees to recover a low-rank
matrix despite gross sparse errors under rather broad
conditions. Mathematically, it considers the matrix $M$ of the
form $M = L_0 + S_0$, where $L_0$ is low-rank and $S_0$ is a
sparse matrix with most of its entries being zero. Unlike the
model for PCA, here both components can be of arbitrary
magnitude and no other information about the rank of $L_0$
and/or the support or signs of $S_0$ is given. To recover $L_0$
and $S_0$, PCP solves the following convex optimization
problem\footnote{In this paper, we use five norms of a matrix
$A$. $\|A\|_*$ denotes its nuclear norm -- sum of its singular
values, $\|A\|_F$ denotes its Frobenius norm and $\|A\|$
denotes its 2-norm. Moreover, $\|A\|_1$ and $\|A\|_{\infty}$
are the $\ell_1$ and $\ell_{\infty}$ norms of $A$ viewed as a
vector, respectively.}
\begin{equation}
\min_{L,S} \; \|L\|_* + \lambda \|S\|_1 \quad \textup{subject to} \quad M
= L + S\label{eqn-rpca}.
\end{equation}
It has been shown in \cite{Wright2009-pp}, under surprisingly
broad conditions, the above convex program exactly recovers
$L_0$ and $S_0$. Readers are also referred to \cite{Chandr09}
which proposed to solve the same problem but with different
exact recovery conditions.

\paragraph{Main Assumptions}
Since our analysis and result will be largely based on the same
conditions of PCP, for completeness, we summarize the precise
conditions and result of PCP here. Let $L_0=U\Sigma
V^*=\sum_{i=1}^r \sigma_i u_i v_i^*$ denote the singular value
decomposition of $L_0\in \R^{n_1\times n_2}$, where $r$ is the
rank, $\sigma_1,\ldots,\sigma_r$ are the singular values, and
$U = [u_1,\ldots,u_r], V=[v_1,\ldots,v_r]$ are the matrices of
left- and right-singular vectors, respectively. The incoherence
conditions on $U$ and $V$ with parameter $\mu$ are as follows:
\begin{small}
\begin{equation}
\max_i\|U^*e_i\|^2\leq\frac{\mu r}{n_1}, \; \max_i\|V^*e_i\|^2\leq\frac{\mu r}{n_2},
\;\|UV^*\|_{\infty}\leq \sqrt{\frac{\mu r}{n_1n_2}},
\label{cond-inc}
\end{equation}
\end{small}where $e_i$'s are the canonical basis vectors. Now
let $\|S_0\|_0 = m$ be the number of nonzero entries in $S_0$.
The conditions on $S_0$ concern the identifiability issue
arises when $S_0$ is also low-rank. To avoid such pathological
cases, \cite{Wright2009-pp} assumes that the support of sparse
component $S_0$ is selected uniformly at random among all
subsets of size $m$. Under these conditions, the main result of
\cite{Wright2009-pp} states:
\begin{theorem}[\cite{Wright2009-pp}]
Suppose $L_0\in \R^{n\times n}$ obeys \eqref{cond-inc} and that
the support set of $S_0$ is uniformly distributed. Then there
is a numerical constant $c$ such that with probability at least
$1-cn^{-10}$ (over the choice of support of $S_0$), Principal
Component Pursuit \eqref{eqn-rpca} with $\lambda=1/\sqrt{n}$
recovers $L_0$ and $S_0$ exactly, provided that
\begin{equation} \label{eqn:assumptions}
\textup{rank}(L_0)\leq \rho_r n \mu^{-1}(\log n)^{-2} \quad and \quad m\leq \rho_s n^2,
\end{equation}
where $\rho_r$ and $\rho_s$ are some positive constants.
\label{thm-rpca}
\end{theorem}
The analysis and result of PCP apply to any rectangular
$(n_1\times n_2)$ matrix, so will be the result of this paper.
But to simplify presentation, we have assumed that the matrices
are all square and write $n=n_1=n_2$. The modification needed
for general rectangular matrices is straightforward and will be
briefly discussed in the end of the paper.

\subsection{Main Result of This Paper}
The PCP result \cite{Wright2009-pp} is limited to the low-rank
component being exactly low-rank and the sparse component being
exactly sparse. However, in real world applications the
observations are often corrupted by noise, which may be
stochastic or deterministic, affecting every entry of the data
matrix. For example, in face recognition, the human face is not
a strictly convex and Lambertian surface hence small
perturbation accounting for the fact that the low-rank
component is only approximately low-rank needs to be
considered. In ranking and collaborative filtering, user's
ratings could be noisy because of the lack of control in the
data collection process. Therefore, for the techniques
developed in \cite{Wright2009-pp} to be widely applicable,
results that guarantee stable and accurate recovery in the
presence of entry-wise noise must be established.

The new measurement model that we consider in this paper
assumes that we observe
\begin{equation}
M = L_0 + S_0 + Z_0,
\end{equation}
where $Z_0$ is a noise term -- say i.i.d. noise on each entry
of the matrix.  However, all we assume about $Z_0$ in this
paper is that $\|Z_0\|_F \leq \delta$ for some $\delta > 0$. To
recover the unknown matrices $L_0$ and $S_0$, we propose
solving the following optimization problem, as a relaxed
version to PCP \eqref{eqn-rpca}:
\begin{equation}
\min_{L,S} \; \|L\|_* + \lambda \|S\|_1 \quad \textup{subject to} \quad \|M
- L - S\|_F \leq \delta. \label{eqn-nrpca}
\end{equation}
where we choose $\lambda = 1/\sqrt{n}$. Our main result is that
under the same conditions as PCP, the above convex program
gives a stable estimate of $L_0$ and $S_0$:
\begin{theorem}
Suppose again that $L_0$ obeys \eqref{cond-inc} and the support
of $S_0$ is uniformly distributed. Then if $L_0$ and $S_0$
satisfy \eqref{eqn:assumptions} with $\rho_r, \rho_s
> 0$ being sufficiently small numerical constants, with high
probability in the support of $S_0$, for any $Z_0$ with $\| Z_0
\|_F \le \delta$, the solution $(\hat{L},\hat{S})$ to the
convex program \eqref{eqn-nrpca} satisfies
\begin{equation}
\| \hat L - L_0 \|_F^2 + \| \hat S - S_0 \|_F^2 \le C n^2 \delta^2,
\label{eqn-mainres}
\end{equation}
where $C$ is a numerical constant. \label{result-main}
\end{theorem}

The precise form of the constant $C$ will be given in
Proposition \ref{prop:stability}. Here, we would like to point
out two ways to view the significance of this result. To some
extent, our model unifies the classical PCA and the robust PCA
by considering both gross sparse errors and small entry-wise
noise in the measurements. So on one hand, our result says that
the low-rank and sparse decomposition via PCP is stable in the
presence of small entry-wise noise, hence making PCP more
widely applicable to practical problems where the low-rank
structure is not exact. On the other hand, together with the
result of PCP \cite{Wright2009-pp}, our new result convincingly
justifies that the classical PCA can now be made robust to
sparse gross corruptions via certain convex programs. Since
this convex program can be solved very efficiently
\cite{Lin09}, at a cost not so much higher than the classical
PCA, our result is expected to have significant impact on many
practical problems.

\subsection{Relations to Existing Work}
Aside from its close relations to the classical PCA and the
newly proposed robust PCA work mentioned above, our analysis
and result are closely related to two lines of development,
regarding stable recovery of sparse signals and low-rank
matrices, respectively.

Conceptually, our work is very similar to the development of
results for the ``imperfect'' scenarios in compressive sensing
where the measurements are noisy and the signal is not exact
sparse. More precisely, $\ell_1$-norm minimization techniques
are adapted to recover a vector $x_0\in \R^m$ from incomplete
and contaminated observations $y=Ax_0+z$ where $A$ is a
$n\times m$ matrix with $n\ll m$ and $z$ is the noise term.
After the landmark work of \cite{CanTao05} which established
that for the noise free case,  the minimal $\ell_1$-norm
solution exactly recovers the sparse signal under fairly broad
conditions, later works have demonstrated that stable recovery
occurs for most measurement ensembles \cite{Donoho04}, or
particularly, when the measurement ensembles satisfy some
simple incoherence conditions \cite{Donoho06} or restricted
isometry property (RIP) \cite{CanRomTao05}.

Recently, there has been an explosion of literature regarding
the power of nuclear-norm minimization in recovering low-rank
matrices from under-sampled measurements. A matrix RIP is first
proposed by \cite{recht-2007} to connect compressive sensing
with low-rank matrix recovery. For measurement ensembles
obeying the RIP, tight bounds of the recovery error from noisy
data have been developed in \cite{CanPla09} which is within a
constant of the minimax risk and an oracle error. Also see
\cite{Negahban09} for similar results. Technically, our work is
more closely related to the recent work \cite{Candes2009-pp}
which developed the first stability result for the matrix
completion problem under small perturbations. Naturally, in
establishing the stability result for robust PCA, we borrow
heavily from the techniques used in \cite{Candes2009-pp} and
\cite{Wright2009-pp}.

\section{Notation and Outline of Analysis}

Our goal is to show that in cases where the noise free
principal component pursuit \eqref{eqn-rpca} {\em exactly}
recovers $(L_0,S_0)$, the noise aware version \eqref{eqn-nrpca}
{\em stably} estimates $(L_0,S_0)$. In the noise free case,
exact recovery is guaranteed by the existence of a ``dual
certificate'' $W$ described in Lemma \ref{lem:duality} below.
The main result of \cite{Wright2009-pp} is to show that under
the stated conditions, with high probability such a dual
certificate exists.  Then Proposition \ref{prop:stability}
below shows that the existence of such a certificate actually
also implies that the recovery via \eqref{eqn-nrpca} under
noise is stable.

Before continuing, we fix some notation. Given a matrix pair
$X_0 = (L_0,S_0)$, let $\Omega \subseteq [n] \times [n]$ denote
the support of $S_0$, and $\P_\Omega$ denote the projection
operator onto the space of matrices supported on $\Omega$. Let
$r = \mathrm{rank}(L_0)$, and let $L_0=U\Sigma V^*$ denote the
compact singular value decomposition of $L_0$, with $U,V \in
\Re^{n\times r}$ and $\Sigma \in \Re^{r \times r}$. We will let
$T$ denote the subspace generated by matrices with the same
column space or row space as $L_0$:
\begin{displaymath}
T = \{UQ^* + RV^* \mid Q,R\in \R^{n\times r}\} \subset \R^{n\times n},
\end{displaymath}
and $\P_T$ be the projection operator onto this subspace.

For any pair $X=(L,S)$ let $\|X\|_F \doteq
(\|L\|_F^2+\|S\|_F^2)^{1/2}$, and define the projection
operator $\P_T \times \P_\Omega : (L,S)\mapsto
(\P_TL,\P_{\Omega}S)$. Define the subspaces $\Gamma \doteq
\{(Q,Q) \mid Q \in \R^{n \times n}\}$ and $\Gamma^\perp \doteq
\{ (Q,-Q) \mid Q \in \R^{n \times n} \}$, and let $\P_\Gamma$
and $\P_{\Gamma^\perp}$ denote their respective projection
operators. Finally, for any linear operator $\mathcal{A} :
\Re^{n \times n} \to \Re^{n \times n}$, we use $\|\mathcal
{A}\|$ to denote the operator norm $\sup_{\|X\|_F=1}\|\mathcal
{A}X\|_F$.

With these notations, the optimality conditions for
\eqref{eqn-rpca} can be stated in terms of a dual vector as
follows.
\begin{lemma}[Lemma 2.5 in \cite{Wright2009-pp}] \label{lem:duality}
Assume that $\|\P_{\Omega}\P_T\| \leq 1/2$ and $\lambda < 1$.
Suppose that there exists $W$ such that
\begin{equation}
\left\{ \begin{array}{l}
W\in T^{\perp},\quad \|W\| < 1/2,\\
\|\P_{\Omega}(UV^*-\lambda\textup{sgn}(S_0)+W)\|_F \leq \lambda/4, \\
\|\P_{\Omega^{\perp}}(UV^*+W)\|_{\infty}<\lambda/2.
\end{array} \right.
\label{cond-weak}
\end{equation}
Then the pair $(L_0,S_0)$ is the unique optimal solution to
\eqref{eqn-rpca}.
\end{lemma}
From now on, we will write $\lambda \P_{\Omega}D =
\P_{\Omega}(UV^*-\lambda\textup{sgn}(S_0)+W)$. The following
proposition shows that under the existence of such a dual
certificate, \eqref{eqn-nrpca} will also stably recover $L_0$
and $S_0$ in the presence of noise.

\begin{prop} \label{prop:stability}
Assume $\|\P_{\Omega}\P_T\|\leq 1/2$, $\lambda\leq 1/2$, and that
there exists a dual certificate $W$ satisfying
\eqref{cond-weak}. Let
$\hat{X}=(\hat{L},\hat{S})$ be the solution to
\eqref{eqn-nrpca} and $X_0 = (L_0,S_0)$, then $\hat{X}$
satisfies
\begin{equation}
\|X_0-\hat{X}\|_F \leq (8\sqrt{5}n  + \sqrt{2})\delta.
\label{eq-mainres}
\end{equation}
\end{prop}

Proposition \ref{prop:stability} implies Theorem
\ref{result-main}, since under the conditions of Theorem
\ref{result-main}, Lemma 2.8 and Lemma 2.9 of
\cite{Wright2009-pp} show that with high probability, there
indeed exists such a dual certificate $W$, and Corollary 2.7 of
\cite{Wright2009-pp} proves $\| \P_\Omega \P_T \| \le 1/2$ as
well.

The rest of the paper then sets out to prove Proposition
\ref{prop:stability} and is organized as follows. In Section
\ref{sec-3}, we prove two key lemmas on which our main result
depends. The proof of Proposition \ref{prop:stability} then
follows in Section \ref{sec-4}. We further provide numerical
results in Section \ref{sec-5} to support our analysis and
conclude the paper with additional discussions in Section
\ref{sec-6}.

\section{Two Lemmas}\label{sec-3}

In this section, we prove two lemmas which will be useful in
the development of our main result. For any matrix pair
$X=(L,S)$, we define $\|X\|_{\diamondsuit} = \|L\|_* +
\lambda\|S\|_1$.
\begin{lemma}
Assume $\|\P_{\Omega}\P_T\|\leq 1/2$ and $\lambda\leq 1/2$.
Suppose that there exists a dual certificate $W$ satisfying
\eqref{cond-weak} and write $\Lambda = UV^* + W$. Then for any
perturbation $H=(H_L,H_S)$ obeying $H_L + H_S = 0$,
\begin{eqnarray*}
\|X_0+H\|_{\diamondsuit} & \geq & \|X_0\|_{\diamondsuit} +
(3/4-\|\P_{T^{\perp}}(\Lambda)\|)\|\P_{T^{\perp}}(H_L)\|_* \\
& & +(3\lambda/4-\|\P_{\Omega^{\perp}}(\Lambda)\|_{\infty})\|\P_{\Omega^{\perp}}(H_S)\|_1.
\end{eqnarray*}
\end{lemma}
\begin{proof}
For any $Z=(Z_L,Z_S)\in \partial \|X_0\|_{\diamondsuit}$, we
have
\begin{displaymath}
\|X_0+H\|_{\diamondsuit} \geq \|X_0\|_{\diamondsuit} + \langle
Z_L,H_L\rangle + \langle Z_S,H_S\rangle .
\end{displaymath}
Now due to the form of the subgradients of the $\ell_1$ norm
and the nuclear norm,\footnote{That is, $Z_S =
\lambda(\textup{sgn}(S_0) + F)$ with $\P_{\Omega} F = 0$ and
$\|F\|_\infty \le 1$; and $Z_L = UV^* + W'$ with $P_T W' = 0$
and $\|W'\| \le 1$.} we have the identities: $Z_L = \Lambda +
\P_{T^{\perp}}(Z_L - \Lambda)$ and $Z_S = \Lambda -
\lambda\P_{\Omega}D + \P_{\Omega^{\perp}}(Z_S - \Lambda)$. Thus
we have:
\begin{eqnarray*}
& & \langle Z_L,H_L\rangle + \langle Z_S,H_S\rangle \\
& = & \langle \Lambda,H_L\rangle + \langle \P_{T^{\perp}}(Z_L - \Lambda),H_L\rangle\\
& & + \langle \Lambda- \lambda\P_{\Omega}D,H_S\rangle + \langle \P_{\Omega^{\perp}}(Z_S - \Lambda),H_S\rangle\\
& \geq & \langle Z_L - \Lambda,\P_{T^{\perp}}(H_L)\rangle \\
& & +\langle Z_S - \Lambda,\P_{\Omega^{\perp}}(H_S)\rangle - \frac{\lambda}{4}\|\P_{\Omega}(H_S)\|_F
\end{eqnarray*}
since $H_L + H_S = 0$ and $\|\P_{\Omega}D\|_F\leq 1/4$.

Moreover, by duality, there exists $Z_L^* \in
\partial\|L_0\|_*$ with $\|Z_L^*\|\leq 1$ such that $\langle
Z_L^*,\P_{T^{\perp}}(H_L)\rangle = \|\P_{T^{\perp}}(H_L)\|_*$.
Also notice that $|\langle \Lambda,\P_{T^{\perp}}(H_L)\rangle |
= |\langle \P_{T^{\perp}}(\Lambda),\P_{T^{\perp}}(H_L)\rangle |
\leq \|\P_{T^{\perp}}(\Lambda)\|\|\P_{T^{\perp}}(H_L)\|_*$.
Therefore, let $Z_L = Z_L^*$, we have:
\begin{equation*}
\langle Z_L - \Lambda,\P_{T^{\perp}}(H_L)\rangle \geq (1-\|\P_{T^{\perp}}(\Lambda)\|)\|\P_{T^{\perp}}(H_L)\|_*.
\end{equation*}
Similarly,  by duality, there
exists $Z_S^* \in \partial (\lambda \|S_0\|_1)$ with $\|Z_S^*\|_{\infty} \leq \lambda$ such that $\langle
Z_S^*,\P_{\Omega^{\perp}}(H_S)\rangle = \lambda
\|\P_{\Omega^{\perp}}(H_S)\|_1$. Therefore, choose $Z_S$ to be $Z_S = Z_S^*$, we have:
\begin{equation*}
\langle Z_S - \Lambda,\P_{\Omega^{\perp}}(H_S)\rangle \geq  (\lambda-\|\P_{\Omega^{\perp}}(\Lambda)\|_{\infty})\|\P_{\Omega^{\perp}}(H_S)\|_1.
\end{equation*}
Observe now that
\begin{eqnarray*}
& & \|\P_{\Omega}(H_S)\|_F
 \leq   \|\P_{\Omega}\P_T(H_S)\|_F + \|\P_{\Omega}\P_{T^{\perp}}(H_S)\|_F \\
& \leq & \frac{1}{2}\|H_S\|_F + \|\P_{T^{\perp}}(H_S)\|_F\\
& \leq & \frac{1}{2}\|\P_{\Omega}(H_S)\|_F + \frac{1}{2}\|\P_{\Omega^{\perp}}(H_S)\|_F + \|\P_{T^{\perp}}(H_S)\|_F,
\end{eqnarray*}
therefore,
\begin{eqnarray*}
\|\P_{\Omega}(H_S)\|_F & \leq & \|\P_{\Omega^{\perp}}(H_S)\|_F + 2\|\P_{T^{\perp}}(H_S)\|_F \\
& \leq & \|\P_{\Omega^{\perp}}(H_S)\|_1 + 2\|\P_{T^{\perp}}(H_L)\|_*.
\end{eqnarray*}
Combining the inequalities above, we have
\begin{eqnarray*}
\|X_0+H\|_{\diamondsuit}
&\geq& \|X_0\|_{\diamondsuit} +
(1-\lambda/2
-\|\P_{T^{\perp}}(\Lambda)\|)\|\P_{T^{\perp}}(H_L)\|_* \\
& & +
(\lambda-\lambda/4 -
\|\P_{\Omega^{\perp}}(\Lambda)\|_{\infty})\|\P_{\Omega^{\perp}}(H_S)\|_1\\
&\geq& \|X_0\|_{\diamondsuit} +
(3/4
-\|\P_{T^{\perp}}(\Lambda)\|)\|\P_{T^{\perp}}(H_L)\|_* \\
& & +
(3\lambda/4-
\|\P_{\Omega^{\perp}}(\Lambda)\|_{\infty})\|\P_{\Omega^{\perp}}(H_S)\|_1.
\end{eqnarray*}
\end{proof}

\begin{lemma}
Suppose that $\|\P_T \P_\Omega \| \le 1/2$. Then for any pair $X = (L,S)$, $\| \P_\Gamma (\P_T \times \P_{\Omega}) (X) \|_F^2 \ge \frac{1}{4} \| (\P_T \times \P_\Omega)(X) \|_F^2.$
\label{lem:operator}
\end{lemma}
\begin{proof}
%
For any matrix pair $X'=(L',S')$,
$\P_{\Gamma}(X') = \Big(\frac{L'+S'}{2},\frac{L'+S'}{2}\Big)$ and so $\| \P_\Gamma(X') \|_F^2 = \tfrac{1}{2} \| L' + S' \|_F^2$. So,
\begin{multline*}
\| \P_\Gamma (\P_T \times \P_\Omega)(X) \|_F^2 = \tfrac{1}{2} \| \P_T(L) + \P_\Omega(S) \|_F^2 \\
= \frac{1}{2} \left( \| \P_T(L) \|_F^2 + \| \P_\Omega(S) \|_F^2 + 2 \< \P_T(L), \P_\Omega(S) \> \right).
\end{multline*}
Now,
\begin{align*}
\< \P_T(L), \P_\Omega(S) \> &= \< \P_T(L), (\P_T \P_\Omega) \P_\Omega (S) \> \\
&\ge - \| \P_T \P_\Omega \| \| \P_T(L) \|_F \| \P_\Omega(S) \|_F.
\end{align*}
Since $\| \P_T \P_\Omega \| \le 1/2$,
\begin{eqnarray*}
\lefteqn{\| P_\Gamma (\P_T\times \P_\Omega)(X) \|_F^2} \\
&\ge& \tfrac{1}{2}\left( \| \P_T(L)\|_F^2 + \| \P_\Omega(S) \|_F^2 - \| P_T(L)\|_F \| P_\Omega(S) \|_F \right) \\
&\ge& \tfrac{1}{4}\left( \| \P_T(L)\|_F^2 + \| \P_\Omega(S) \|_F^2 \right) = \tfrac{1}{4}\| (\P_T \times \P_\Omega)(X) \|_F^2,
\end{eqnarray*}
where we have used that for any $a,b$, $a^2 + b^2 - ab \ge (a^2
+ b^2)/2$.
\end{proof}

\section{Proof of Proposition \ref{prop:stability}} \label{sec-4}

Our proof uses two crucial properties of $\hat{X}$. First,
since $X_0$ is also a feasible solution to \eqref{eqn-nrpca},
we have $\|\hat{X}\|_{\diamondsuit} \leq
\|X_0\|_{\diamondsuit}$.
Second, we use triangle inequality to get
\begin{eqnarray}
&& \|\hat{L}+\hat{S}-L_0-S_0\|_F \nonumber \\ &\leq& \|\hat{L}+\hat{S}-M\|_F + \|L_0+S_0-M\|_F \leq 2\delta .
\label{eq-tube}
\end{eqnarray}

Furthermore, set $\hat{X} = X_0 + H$ where $H=(H_L,H_S)$ and
write $H^{\Gamma}=\P_{\Gamma}(H)$,
$H^{\Gamma^{\perp}}=\P_{\Gamma^{\perp}}(H)$ for short. We want
to bound $\|H\|_F^2$, which can be expanded as
\begin{align}
& \|H\|_F^2 = \|H^{\Gamma}\|_F^2 +
\|H^{\Gamma^{\perp}}\|_F^2 \nonumber\\
& = \|H^{\Gamma}\|_F^2 + \|(\P_{T}\times \P_{\Omega})(H^{\Gamma^{\perp}})\|_F^2 + \|(\P_{T^{\perp}}\times \P_{\Omega^{\perp}})(H^{\Gamma^{\perp}})\|_F^2.
\label{eq-part2}
\end{align}
Since \eqref{eq-tube} gives us $\|H^{\Gamma}\|_F  = \big
(\|(H_L+H_S)/2\|_F^2 + \|(H_L+H_S)/2\|_F^2\big )^{1/2} \leq
\sqrt{2}/2 \times 2\delta = \sqrt{2}\delta$, it suffices to
bound the second and third terms on the right-hand-side of
\eqref{eq-part2}.

{\bf a. Bound the third term of \eqref{eq-part2}.} Let $W$ be a
dual certificate satisfying \eqref{cond-weak}. Then, $\Lambda =
UV^*+W$ obeys $\|\P_{T^{\perp}}(\Lambda)\| \leq 1/2$ and
$\|\P_{\Omega^{\perp}}(\Lambda)\|_{\infty} \leq \lambda/2$. We
have
\begin{equation}
\|X_0+H\|_{\diamondsuit} \geq \|X_0+H^{\Gamma^{\perp}}\|_{\diamondsuit} - \|H^{\Gamma}\|_{\diamondsuit}
\end{equation}
and
\begin{eqnarray*}
& & \|X_0+H^{\Gamma^{\perp}}\|_{\diamondsuit} \\
& \geq & \|X_0\|_{\diamondsuit} + (3/4-\|\P_{T^{\perp}}(\Lambda)\|)\|\P_{T^{\perp}}(H^{\Gamma^{\perp}}_L)\|_* \\
& & + \big(3\lambda/4-\|\P_{\Omega^{\perp}}(\Lambda)\|_{\infty}\big)\|\P_{\Omega^{\perp}}(H^{\Gamma^{\perp}}_S)\|_1\\
& \geq & \|X_0\|_{\diamondsuit} + \frac{1}{4}\Big( \|\P_{T^{\perp}}(H^{\Gamma^{\perp}}_L)\|_* + \lambda \|\P_{\Omega^{\perp}}(H^{\Gamma^{\perp}}_S)\|_1\Big),
\end{eqnarray*}
which implies that
\begin{equation}
\|\P_{T^{\perp}}(H^{\Gamma^{\perp}}_L)\|_* + \lambda \|\P_{\Omega^{\perp}}(H^{\Gamma^{\perp}}_S)\|_1 \leq 4\|H^{\Gamma}\|_{\diamondsuit}.
\end{equation}
For any matrix $Y \in \R^{n\times n}$, we have the following
inequalities:
\begin{eqnarray*}
\|Y\|_F \leq \|Y\|_* \leq \sqrt{n}\|Y\|_F,
\frac{1}{\sqrt{n}}\|Y\|_F \leq \lambda \|Y\|_1 \leq \sqrt{n}\|Y\|_F,
\end{eqnarray*}
where we assume $\lambda = \frac{1}{\sqrt{n}}$. Therefore
\begin{eqnarray}
& & \|(\P_{T^{\perp}}\times \P_{\Omega^{\perp}})(H^{\Gamma^{\perp}})\|_F \nonumber \\
& \leq & \|\P_{T^{\perp}}(H^{\Gamma^{\perp}}_L)\|_F + \|\P_{\Omega^{\perp}}(H^{\Gamma^{\perp}}_S)\|_F \nonumber \\
& \leq & \|\P_{T^{\perp}}(H^{\Gamma^{\perp}}_L)\|_* + \lambda \sqrt{n} \|\P_{\Omega^{\perp}}(H^{\Gamma^{\perp}}_S)\|_1 \label{eq-t1}\nonumber \\
& \leq & 4\sqrt{n}\|H^{\Gamma}\|_{\diamondsuit} \label{eq-t2}
=  4 \sqrt{n} (\|H^{\Gamma}_L\|_* + \lambda \|H^{\Gamma}_S\|_1) \nonumber \\
& \leq & 4n(\|H^{\Gamma}_L\|_F + \|H^{\Gamma}_S\|_F) \label{eq-t3}
=  4\sqrt{2}n\|H^{\Gamma}\|_F \leq 8n\delta,
\label{eq-part2a}
\end{eqnarray}
where the last equation uses the fact that $H^{\Gamma}_L =
H^{\Gamma}_S$.

{\bf b. Bound the second term of \eqref{eq-part2}.}
By Lemma \ref{lem:operator},
\begin{eqnarray*}
\|\P_{\Gamma}(\P_{T}\times \P_{\Omega})(H^{\Gamma^{\perp}})\|_F^2 \ge \frac{1}{4} \| (\P_T\times \P_\Omega) (H^{\Gamma^\perp}) \|_F^2.
\end{eqnarray*}
But since $\P_{\Gamma}(H^{\Gamma^{\perp}}) = 0 =
\P_{\Gamma}(\P_{T}\times \P_{\Omega})(H^{\Gamma^{\perp}}) +
\P_{\Gamma}(\P_{T^{\perp}}\times
\P_{\Omega^{\perp}})(H^{\Gamma^{\perp}})$, we have
\begin{eqnarray*}
\|\P_{\Gamma}(\P_{T}\times \P_{\Omega})(H^{\Gamma^{\perp}})\|_F  & = & \|\P_{\Gamma}(\P_{T^{\perp}}\times \P_{\Omega^{\perp}})(H^{\Gamma^{\perp}})\|_F \\
& \leq & \|(\P_{T^{\perp}}\times \P_{\Omega^{\perp}})(H^{\Gamma^{\perp}})\|_F.
\end{eqnarray*}
Combining the previous two inequalities, we have
\begin{eqnarray*}
\|(\P_{T}\times \P_{\Omega})(H^{\Gamma^{\perp}})\|_F^2 & \leq & 4 \|(\P_{T^{\perp}}\times \P_{\Omega^{\perp}})(H^{\Gamma^{\perp}})\|_F^2,
\label{eq-part2b}
\end{eqnarray*}
which, together with \eqref{eq-part2a}, gives us the desired
result,
\begin{equation}
\|H^{\Gamma^{\perp}}\|_F^2 \leq 5\|(\P_{T^{\perp}}\times \P_{\Omega^{\perp}})(H^{\Gamma^{\perp}})\|_F^2 \leq 64\times 5 \times n^2\delta^2.
\end{equation}

\section{Simulations} \label{sec-5}
In this section, we run a series of numerical experiments on
square matrices with noisy entries. For each setting of
parameters, we report the average errors over 20 trials. Each
entry of the noise term ${Z_0}$ is i.i.d. $N(0,\sigma^2)$. A
rank-$r$ matrix $L_0$ is generated as $L_0 = UV^*$ where both
$U$ and $V$ are $n\times r$ matrices with i.i.d.
$N(0,\sigma_n^2)$ entries, with $\sigma_n^2 \doteq
10\frac{\sigma}{\sqrt{n}}$. Here, the value of $\sigma_n$ is
rather arbitrary and set such that the singular values of $L_0$
are much larger than the singular values of $Z_0$. The entries
of $S_0$ are independently distributed, each taking on value
$0$ with probability $1-\rho_s$, and uniformly distributed in
$[-5,5]$ with probability $\rho_s$.

In order to stably recover $\hat{X}=(\hat{L},\hat{S})$, instead
of directly solving \eqref{eqn-nrpca}, we solve the following
dual problem, to which a fast proximal gradient algorithm
proposed in \cite{Lin09}, \emph{Accelerated Proximal Gradient}
(APG), can be applied.
\begin{equation}
\min_{L,S} \; \|L\|_* + \lambda \|S\|_1 + \frac{1}{2\mu} \|M
- L - S\|_F^2 \label{eqn-dual}.
\end{equation}
It is well established that \eqref{eqn-dual} is equivalent to
\eqref{eqn-nrpca} for some value $\mu(\delta)$. Our choice of
$\mu$ here follows similar arguments as in
\cite{Candes2009-pp}. First, note that if we fix $S=0$ in
\eqref{eqn-dual}, the solution $\hat{L}$ of \eqref{eqn-dual} is
equal to the singular value thresholding version of $M$ with
threshold $\mu$. Similarly, if we fix $L=0$ in
\eqref{eqn-dual}, the solution $\hat{S}$ is equal to the
entry-wise shrinkage version of $M$ with threshold $\mu
\lambda$. Thus, we choose $\mu$ to be the smallest value such
that the minimizer of \eqref{eqn-dual} is likely to be
$\hat{L}=\hat{S}=0$ if we set $L_0=S_0=0$ and $M=Z_0$. In this
way, $\mu$ is large enough to threshold away the noise, but not
too large to over-shrink the original matrices. Now, it is well
known that for $Z_0\in \R^{n\times n}$, $n^{-1/2}\|Z_0\|
\rightarrow \sqrt{2}\sigma$ almost surely as $n \rightarrow
\infty$. Thus, we choose $\mu=\sqrt{2n}\sigma$. This also fits
the sparse component well since $\mu\lambda = \sqrt{2}\sigma$.
We shall see that this choice of $\mu$ works well in practice.

\subsection{Comparison with An Oracle}
To further understand our algorithm, we would like to compare
its performance to the best possible accuracy one can achieve,
for instance, by the minimal mean-square-error (MMSE) estimator
over all low-rank and sparse matrix pairs. However, because
obtaining the MMSE estimation is not computationally tractable,
we instead resort to an oracle which gives us information about
the support $\Omega$ of $S_0$ and the row and column spaces $T$
of $L_0$. Our oracle estimates $L$ and $S$ as the solution
$L_{oracle}$ and $S_{oracle}$ to the following least squares
problem:
\begin{equation}
\min_{L,S} \; \|M - L - S\|_F \quad \textup{subject to} \quad
L \in T, S\in \Omega.
\label{eqn-oracle}
\end{equation}
Since we know the locations of the corrupted entries, we can
solve for $L_{oracle}$ and $S_{oracle}$ separately. That is, we
first find the matrix in $T$ which best fits the uncorrupted
data in a least squares sense. Under the hypotheses of Theorem
\ref{prop:stability}, the operator
$\P_T\P_{\Omega^{\perp}}\P_T$ is invertible\footnote{In fact,
since $\|\P_T\P_{\Omega}\P_T\| = \|\P_{\Omega}\P_T\|^2 \leq
1/4$, the smallest eigenvalue of $\P_T\P_{\Omega^{\perp}}\P_T$
is bounded below by $1-1/4=3/4$.} when restricted to $T$ and
the least squares solution is given by
\begin{eqnarray*}
L_{oracle} 
& = & (\P_T\P_{\Omega^{\perp}}\P_T)^{-1}\P_T\P_{\Omega^{\perp}}(M).
\end{eqnarray*}
and the sparse component is given by \begin{equation*}
S_{oracle} = \P_{\Omega}(M - L_{oracle}).
\end{equation*}

\subsection{Experiment Results and Analysis}

\begin{figure}[t]
\centering
\begin{tabular}{cc}
\hspace{-.2in}\includegraphics[scale=0.35,clip=true]{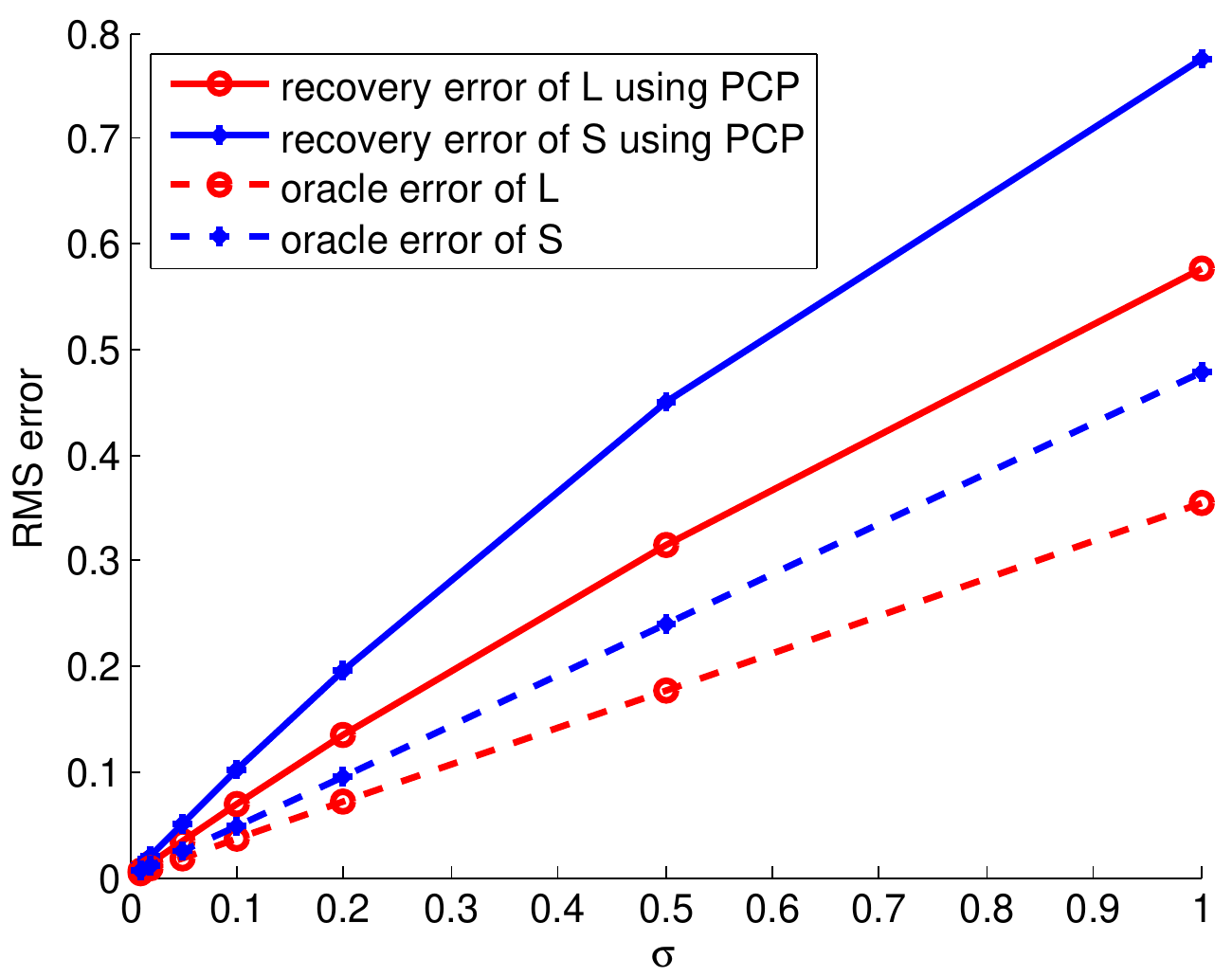} \hspace{-.2in} &
\includegraphics[scale=0.35,clip=true]{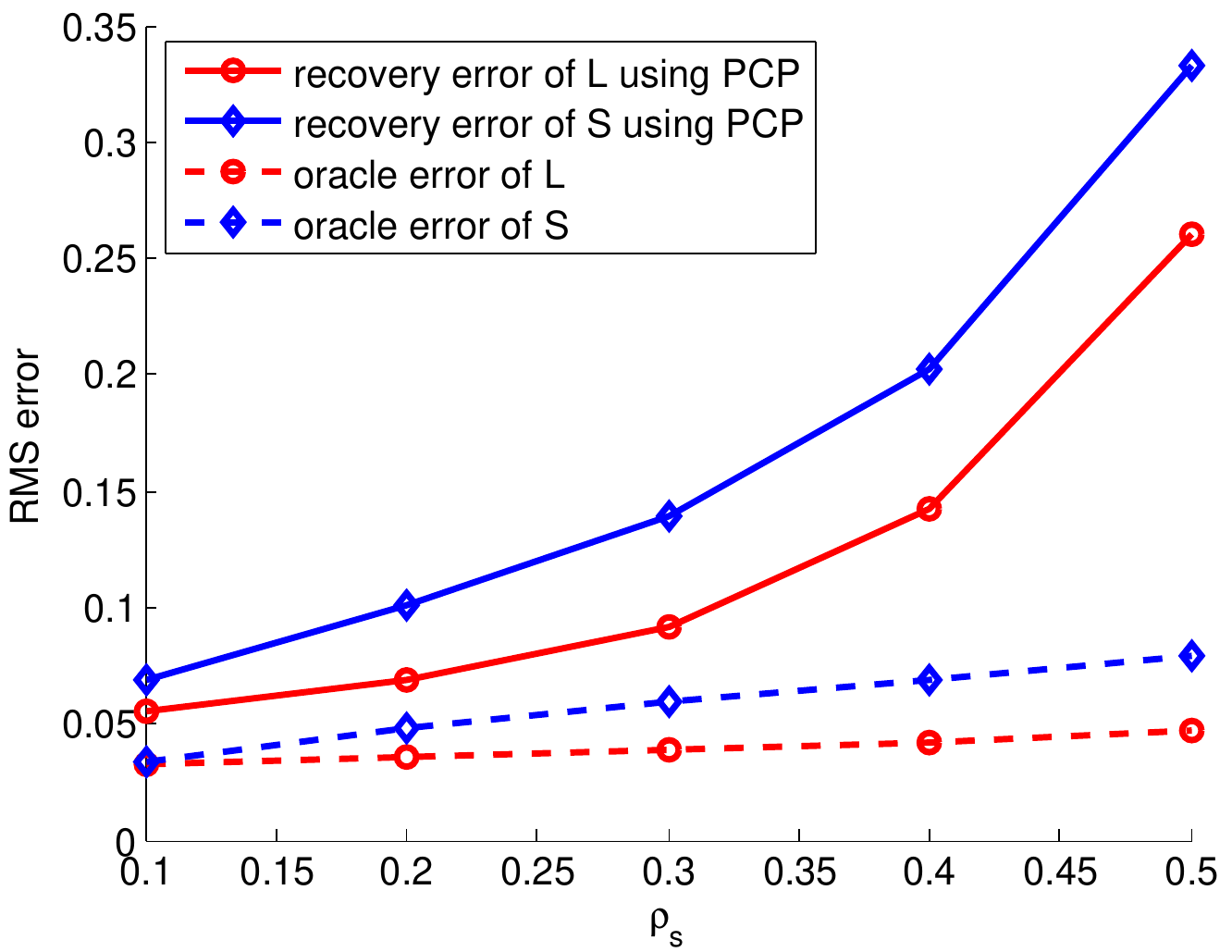}\\
(a) & (b)
\end{tabular}
\caption{(a) RMS errors as a function of $\sigma$ with $r=10, \rho_s = 0.2, n=200$. (b) RMS errors as a function of $\rho_s$ with $r=10, \sigma=0.1, n=200$.}\vspace{-.15in}
\label{fig-1}
\end{figure}

We first evaluate the performance of \eqref{eqn-dual} with
matrix $L_0$ whose rank $r=10$ is fixed. We measure estimation
errors using the root-mean-squared (RMS) error as
$\|\hat{L}-L_0\|_F/n$, $\|\hat{S}-S_0\|_F/n$ for the low-rank
component and the sparse component, respectively.
Fig.~\ref{fig-1}(a) shows the RMS error with varying noise
level $\sigma$. In this experiment, the dimension $n=200$ and
the fraction of corrupted entries $\rho_s=0.2$ are fixed. As
predicted by our main result, the RMS error grows approximately
linearly with the noise level. Moreover, the RMS error by
solving \eqref{eqn-nrpca} is just about twice the RMS error
achieved by the oracle introduced in the previous section.

\begin{figure}[t]
\centering
\begin{tabular}{cc}
\hspace{-.2in}\includegraphics[scale=0.35,clip=true]{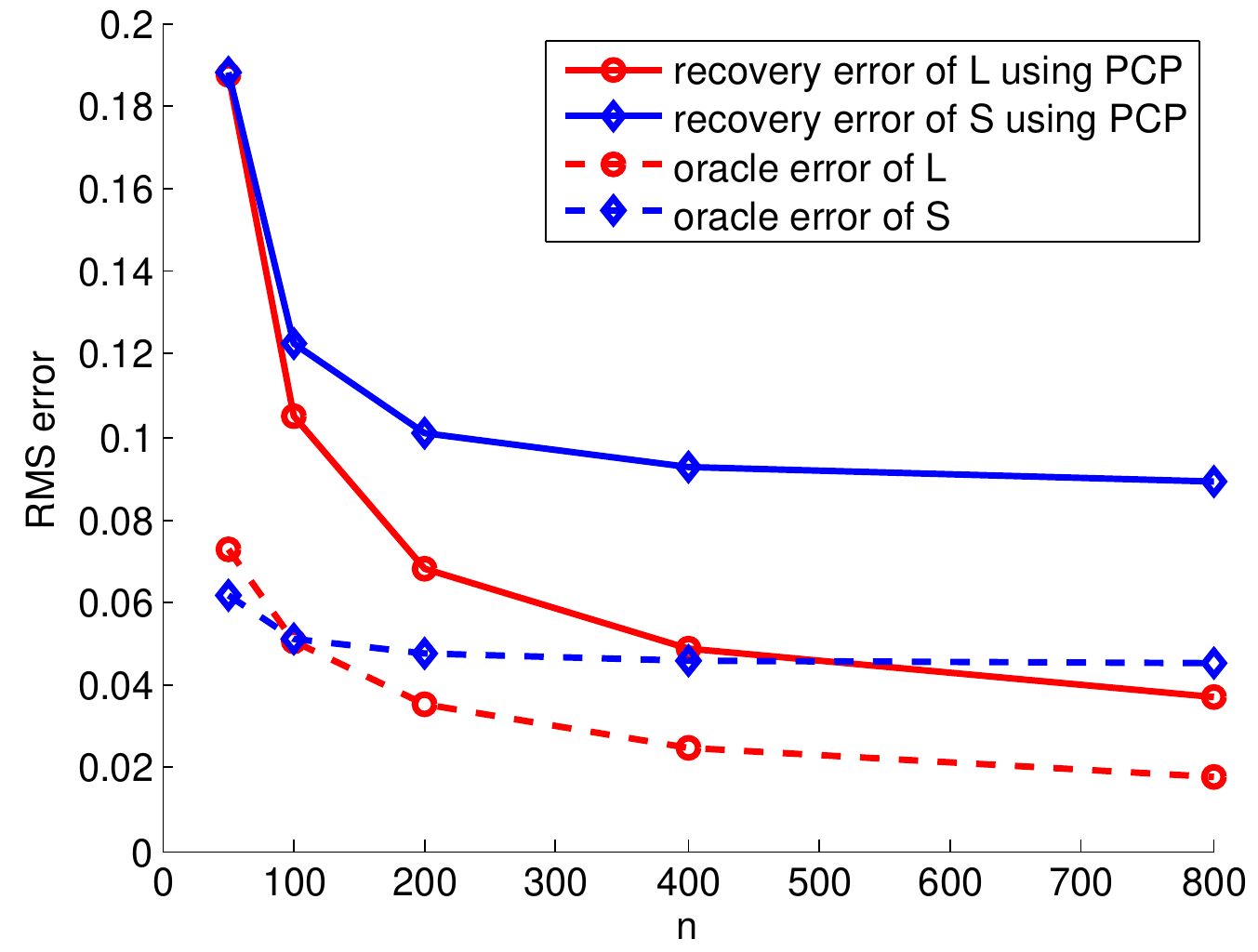} \hspace{-.2in} &
\includegraphics[scale=0.35,clip=true]{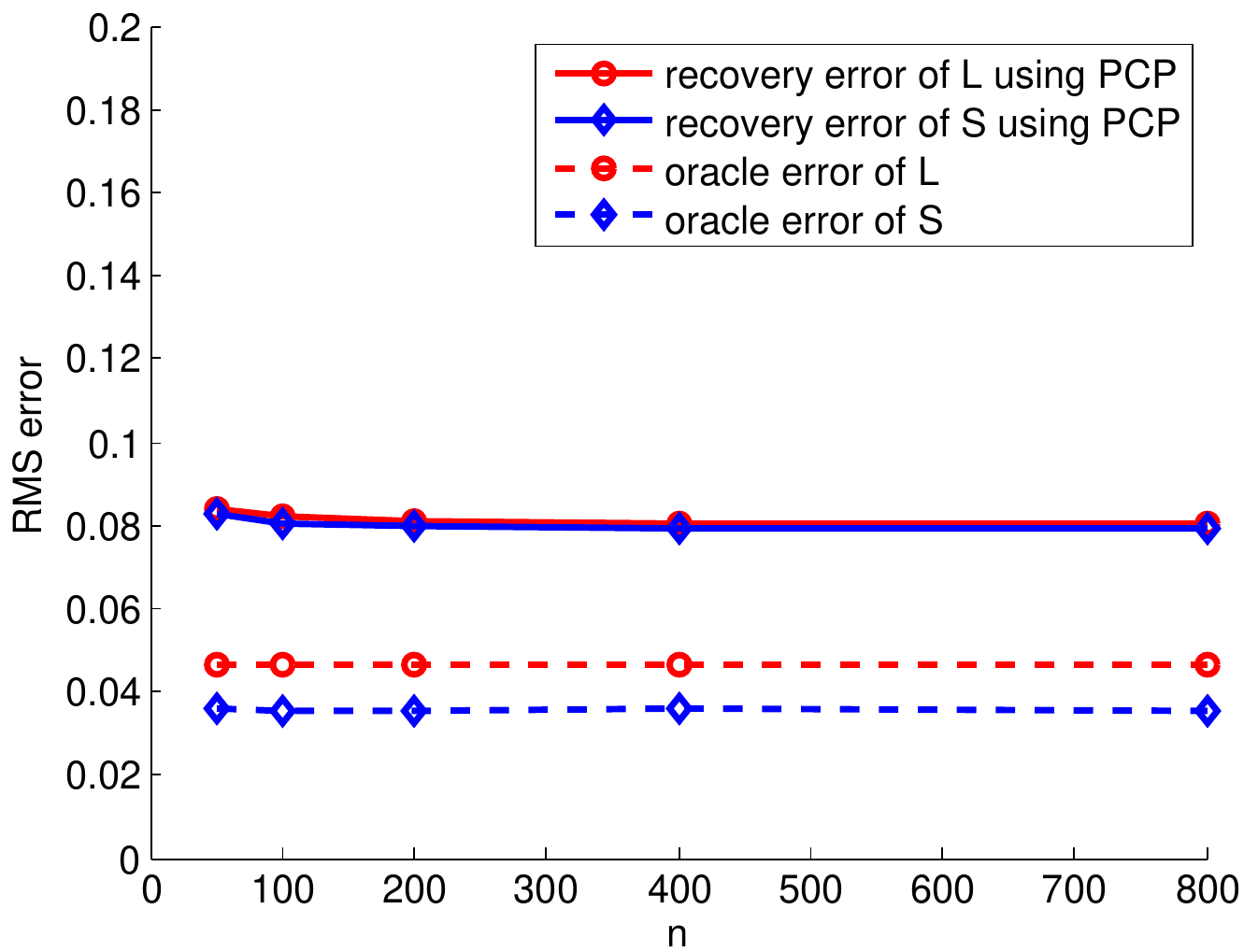}\\
(a) & (b)
\end{tabular}
\caption{RMS errors as a function of $n$ with (a) $\sigma=0.1, \rho_s = 0.2, r=10$ fixed, (b) $\sigma=0.1, \rho_s = 0.1$ and $r = 0.1 \times n$ growing in proportion to $n$.}\vspace{-.15in}
\label{fig-2}
\end{figure}

Now we fix $\sigma=0.1$. Fig.~\ref{fig-1}(b) and
Fig.~\ref{fig-2}(a) show the results with varying $\rho_s$
(when $n=200$ is fixed) and $n$ (when $\rho_s$=0.2 is fixed).
Fig.~\ref{fig-1}(b) illustrates that one can achieve higher
breakdown point by knowing $\Omega$ and $T$. It is observed in
\cite{Wright2009-pp} that when the rank $r$ is fixed or grows
sufficiently slowly as $n$ increases, our method can recover
more and more corrupted entries. Here in Fig.~\ref{fig-2}(a) we
see a similar phenomenon. As $n$ increases, the RMS error
decreases given a fixed fraction of corrupted entries. That is,
our approach can simultaneously tolerate a large fraction of
corrupted entries and a high level of noise when the dimension
$n$ is sufficiently large.


To further test the stability of \eqref{eqn-dual}, we examine
how the algorithm performs when the rank of $L_0$ grows in
proportion to $n$ and the fraction of errors in $S_0$ grows in
proportion to $n^2$. More precisely, in Fig.~\ref{fig-2}(b) we
fix $\sigma = 0.1$, and plot the RMS error as a function of
$n$, with $\mathrm{rank}(L_0) = 0.1 \times n$ and $\rho_s =
0.1$. The result clearly shows that our approach can recover a
wide range of matrix pairs $(L_0,S_0)$, in the presence of
noise. Interestingly, these results also suggest that our
analysis loses a factor of $n$ with respect to the optimal
bound.

\section{Discussion} \label{sec-6}

In this paper, we only present the result for square matrices
for simplicity. However, the arguments and results can be
easily modified to handle the general case. For instance, when
the matrices are $n_1\times n_2$, let $n_{(1)}=\max (n_1,n_2)$
and $n_{(2)}=\min (n_1,n_2)$. The conclusion of Theorem
\ref{thm-rpca} can be stated as: PCP with
$\lambda=1/\sqrt{n_{(1)}}$ succeeds with probability at least
$1-cn_{(1)}^{-10}$, provided that $\textup{rank}(L_0)\leq\rho_r
n_{(2)} \mu^{-1}(\log n_{(1)})^{-2}$ and $m\leq \rho_s n_1n_2$.
Also, relation \eqref{eqn-mainres} in Theorem \ref{result-main}
becomes $\| \hat L - L_0 \|_F^2 + \| \hat S - S_0 \|_F^2 \le C
n_1n_2 \delta^2$.

As suggested by the numerical results, one could hope to
improve the stability result by removing the dependence on $n$.
In this direction, we would like to point out that most of our
analysis seems to be tight, except \eqref{eq-part2a} where we
invoke the generic relations between the nuclear norm, $\ell_1$
norm and the Frobenius norm. Fully examination of this problem
may require additional model assumptions. It is also very
likely that some results in the geometry of Banach spaces,
namely the spherical sections theorem and concentration of
measure, will play a key role in it.
\bibliographystyle{IEEEtran}
\bibliography{isit_noise}

\end{document}